\documentclass[12pt]{article}

\usepackage{amssymb}
\usepackage{amsmath}
\usepackage{amscd}
\usepackage{latexsym}

\usepackage{graphicx}
\usepackage{subfig}

\usepackage{url}

\usepackage{cite}
\usepackage{caption}

\newcommand{\be}{\begin{equation}}
\newcommand{\ee}{\end{equation}}

\newcommand{\dlt}{\delta}

\newcommand{\bt}{\beta}
\newcommand{\vp}{\varphi}
\newcommand{\ep}{\varepsilon}
\newcommand{\al}{\alpha}
\newcommand{\ra}{\rightarrow}

\newcommand{\cD}{{\cal D}}

\newcommand{\cH}{{\cal H}}

\newcommand{\rgl}{\rangle}
\newcommand{\lgl}{\langle}

\begin{document}

\begin{center}

{\Large{\bf Evolutional entanglement production} \\ [5mm]

V.I. Yukalov$^{1,*}$ and E.P. Yukalova$^2$ } \\ [3mm]

{\it
$^1$Bogolubov Laboratory of Theoretical Physics, \\
Joint Institute for Nuclear Research, Dubna 141980, Russia \\ [3mm]

$^2$Laboratory of Information Technologies, \\
Joint Institute for Nuclear Research, Dubna 141980, Russia }

\end{center}

\vskip 5cm

\begin{abstract}

Evolutional entanglement production is defined as the amount of
entanglement produced by the evolution operator. This quantity is
analyzed for systems whose Hamiltonians are characterized by spin
operators. The evolutional entanglement production at the initial 
stage grows quadratically in time. For longer times, it oscillates, 
being quasiperiodic or periodic depending on the system parameters.   

\end{abstract}

\vskip 1cm

{\parindent = 0pt
{\bf PACS numbers}: 03.65.Ud, 03.67.Bg

\vskip 2cm

{\bf $^*$corresponding author}: V.I. Yukalov

{\bf E-mail}: yukalov@theor.jinr.ru   }

\newpage

\section{Introduction}

The notion of entanglement is at the center of several interrelated
problems, such as quantum information processing, quantum computing,
quantum measurements, and quantum decision theory
\cite{Williams_1,Nielsen_2,Vedral_3,Keyl_4,Wilde_5,Yukalov_6}. A closely
related notion is {\it entanglement production} that characterizes the
amount of entanglement produced by quantum operations
\cite{Zanardi_7,Zanardi_8,Yukalov_9,Yukalov_10,Yukalov_11,Yukalov_12,
Vedral_13,Martinez_14,Strom_15,Chen_16}.

The difference between the notions of {\it state entanglement} and 
{\it entanglement production} is as follows. The state entanglement
characterizes the structure of a given state. For example, whether the
system state can be represented as a product of partial states or not 
\cite{Williams_1,Nielsen_2,Vedral_3,Keyl_4,Wilde_5}. While entanglement 
production, induced by a quantum operation, shows whether this operation 
transforms a disentangled state to an entangled one or not. To be more 
precise, let us consider a system associated with a Hilbert space 
$\mathcal{H}$ and let the system be decomposable onto parts associated
with Hilbert spaces $\mathcal{H}_i$. Let the system be described by
a disentangled state of the product form
$$
 \psi_{dis} = \bigotimes_i \psi_i \;  ,
$$
in which $\psi_i \in \mathcal{H}_i$. But assume that we need to transform 
the disentangled state into an entangled one. The necessity of transforming 
a disentangled state into an entangled state can be dictated by the desire 
of using the entangled state for some applications in quantum information 
processing or quantum computing, where the use of entangled states is known 
to be essentially more efficient than the use of disentangled states 
\cite{Williams_1,Nielsen_2,Vedral_3,Keyl_4,Wilde_5}. The required 
transformation can be realized by an operation described by an operator,
say $\hat{A}$, such that its action on the given disentangled state yields
an entangled state
$$
   \psi_{ent} = \hat A \psi_{dis}
$$
that cannot be represented in the product form of partial states. 

For a single given disentangled state, it is possible to find an appropriate
operator. However, the typical question, accompanying the process of such 
transformations, is: How would it be possible to find an operation that would 
be the most efficient for entangling, not just one given state, but the states 
from the whole class of disentangled states of the considered Hilbert space? 
To answer this question, it is necessary to have a characteristic quantifying
the ability of different operators to produce entangled states. Such a 
characteristic has been given by the measure of entanglement production of 
quantum operations \cite{Yukalov_9,Yukalov_10}. In these papers, the use of 
the introduced measure was illustrated by numerous cases of pure as well as
mixed states. It was also shown that entanglement production by reduced 
density operators can be employed for characterizing phase transitions in 
statistical systems, so that thermodynamic phase transitions are usually 
accompanied by {\it entanglement transitions}. For instance, in Bose-Einstein
condensation, entanglement production by density operators decreases, which
also happens in paramagnetic to ferromagnetic transition, contrary to the 
increase of the related measure in the transition from normal metal to 
superconductor. In Ref. \cite{Yukalov_11}, it was shown that entanglement 
produced by atomic correlations through the common radiation field 
experiences sharp peaks in the regime of electromagnetic superradiance. 
In Refs. \cite{Yukalov_12,Yukalov_17}, it was demonstrated that entanglement 
can be produced in a Bose-condensed system by an external alternating field 
creating multiple coherent topological modes. The same can be done by shaking 
an optical lattice filled with Bose-Einstein condensate \cite{Yukalov_18}. 
The consequences of entanglement production can be noticed in time-of-flight 
experiments \cite{Yukalov_12}. In Refs. \cite{Yukalov_6,Yukalov_19}, it was 
studied how the process of quantum measurements can produce entanglement in 
a multi-mode quantum system.  

Instead of producing entanglement by some operations, it is possible to
let the given disentangled state naturally evolve in time until it becomes
entangled. Such a process is described by the evolution operator $\hat{U}(t)$,
with the corresponding evolution generator playing the role of the system
Hamiltonain. In this case, a system, starting from an initial non-entangled 
state can become entangled in the process of its natural evolution with the 
given Hamiltonian, so that
$$ 
\psi_{ent}(t) = \hat U(t) \psi_{dis}(0) \;   .
$$
This kind of time-dependent entanglement can be described by a measurement 
procedure accomplished in a sequence of times with calculating, e.g., 
concurrence at these different time moments \cite{Mintert_29}. This method 
gives the sequence of values characterizing the state entanglement at different
times. The efficiency of entanglement production for given initial and final 
states can be associated with the entanglement probability
$$
p_{ent}(t) \equiv |\; ( \psi_{ent}(t) \; | \; \psi_{dis}(0) ) \; |^2 =
|\; ( \psi_{dis}(0) \; | \; \hat U(t) \psi_{dis}(0) ) \; |^2   \; ,
$$
introduced by analogy with the transition probability. The quantum transition 
probabilities of the type
$$
p(\psi_1  \ra \psi_2) \equiv |\; ( \psi_2 \; | \; \psi_1 ) \; |^2
$$
are widely used in numerous applications characterizing different quantum
transitions, return probability, and quantum many-body localization
\cite{Heller_20,Logan_21,Basko_22,Pal_23,Huse_24}. However, such 
probabilities are defined for the given pair of an initial and final states,
being strongly dependent on them. Again, aiming at quantifying the entangling 
properties of the evolution operator, not just for a given pair of states, but
for a whole class of states from the considered Hilbert space, we can employ 
the measure of entanglement production introduced in Refs. 
\cite{Yukalov_9,Yukalov_10}. 

It is the aim of the present paper to define entanglement production caused 
by the evolution operator and to study its temporal behavior for some concrete 
examples. For this illustration, we choose the systems that are characterized 
by Hamiltonians expressed through spin operators. Such a type of Hamiltonians 
is generic for many systems describing finite-level or finite-state physical 
objects. Many finite quantum systems can be approximated by finite-level
models, when only several low-lying energy levels are involved in the studied
physical processes \cite{Birman_30}. The entanglement production by the evolution 
operator has not been considered in the previous papers.

\section{Measure of entanglement production}

Let us consider a system characterized by the Hilbert space
\be
\label{1}
 \cH = \bigotimes_{i=1}^N \cH_i \;  ,
\ee
where each of the spaces $\mathcal{H}_i$ is a closed linear envelope of
an orthonormal basis of microstates,
$$
\cH_i = {\rm  span}\{ | n_i \rgl \} \;   .
$$
Then the basis in space (\ref{1}) is formed by the states
\be
\label{2}
 | n_1 n_2 \ldots n_N \rgl \equiv \bigotimes_{i=1}^N  | n_i \rgl  \;  ,
\ee
so that
\be
\label{3}
 \cH = {\rm  span}\left \{ \bigotimes_{i=1}^N | n_i \rgl \right \} \;  .
\ee

Among the states of space (\ref{3}), it is possible to separate two types
of qualitatively different states, disentangled and entangled. The set of
disentangled states
\be
\label{4}
\cD \equiv \left \{ \vp = \bigotimes_{i=1}^N \vp_i \right \} \subset \cH
\ee
is formed by the states that are represented as factors of the partial
states
$$
\vp_i = \sum_{n_i} c_{n_i} | n_i \rgl \in \cH_i \;   .
$$
The states that cannot be represented as such factor states are called
entangled.

Let us be interested in the action of an operator $\hat{A}$, with a nonzero
trace, acting on space (\ref{3}). Generally, its action on a state
$\varphi \in \mathcal{H}$ can produce an entangled state, even when the
state $\varphi$ is disentangled. The measure of this entanglement production
can be quantified in the following way \cite{Yukalov_9,Yukalov_10}. For a
given operator $\hat{A}$, we define its non-entangling counterpart
\be
\label{5}
 \hat A^{\otimes} \equiv
\frac{\bigotimes_{i=1}^N \hat A_i}{({\rm Tr}_\cH\hat A)^{N-1}} \; ,
\ee
in which a partial factor operator
\be
\label{6}
\hat A_i \equiv {\rm Tr}_{\cH/\cH_i} \hat A
\ee
is obtained by taking the trace of $\hat{A}$ over all spaces $\mathcal{H}_j$,
composing $\mathcal{H}$, except the single space $\mathcal{H}_i$. The
so-defined non-entangling counterpart (\ref{5}) enjoys the same normalization
as $\hat{A}$, so that
\be
\label{7}
 {\rm Tr}_\cH \hat A^{\otimes} = {\rm Tr}_\cH \hat A  \; .
\ee

For what follows, we need the definition of an operator norm. We opt for the
Hilbert-Schmidt norm that for an operator $\hat{A}$ reads as
$$
 || \hat A||_\cH \equiv \sqrt{   {\rm Tr}_\cH (\hat A^+ \hat A ) }
\equiv || \hat A|| \;  .
$$
Respectively, for an operator $\hat{A}_i$ on the space $\mathcal{H}_i$, the
norm is
$$
 || \hat A_i||_{\cH_i} \equiv
\sqrt{   {\rm Tr}_{\cH_i} (\hat A^+_i \hat A_i ) } \equiv || \hat A_i|| \;  .
$$
This norm, also termed the Frobenius norm or Schur norm, is analogous to the 
Euclidean norm for vectors. It is a particular case $(p=2)$ of the Schatten 
$p$-norm, and, as all Schatten norms, it is invariant under unitary 
transformations \cite{Weidmann_25,Bhatia_26}, thus, does not depend on the 
chosen basis. 

The measure of entanglement production for an operator $\hat{A}$ is defined
\cite{Yukalov_9,Yukalov_10} as
\be
\label{8}
 \ep(\hat A) \equiv
\log \; \frac{ || \hat A||}{|| \hat A^{\otimes} || } \;,
\ee
where the logarithm can be taken with respect to any convenient base. This
quantity satisfies all properties required for being considered as a
measure \cite{Yukalov_9,Yukalov_10}. In particular, when the operator
$\hat{A}$ is not entangling, then $\varepsilon(\hat{A}) = 0$.

For the norm of the non-entangling operator (\ref{5}), we have
\be
\label{9}
 || \hat A^{\otimes} ||^2 =
\frac{\prod_{i=1}^N || \hat A_i||^2}{ |\; {\rm Tr}_\cH \hat A\; |^{2(N-1)} } \; .
\ee
Therefore the measure of entanglement production (\ref{8}) can be represented
in the form
\be
\label{10}
 \ep(\hat A) = \log \left \{ || \hat A|| \; |\; {\rm Tr}_\cH \hat A\; |^{N-1}
 \prod_{i=1}^N || \hat A_i||^{-1} \right \} \; .
\ee
It is important that the defined measure (\ref{8}) or (\ref{10}) is very general, 
being introduced for arbitrary operators with non-zero trace and for arbitrary 
systems, whether pure or mixed, bipartite or multipartite. Many examples of its 
application to concrete physical systems can be found in Refs. 
\cite{Yukalov_9,Yukalov_10,Yukalov_11,Yukalov_12,Yukalov_17,Yukalov_18,Yukalov_19}.
More concretely, the results of the previous papers are described in the 
Introduction. We emphasize that the entanglement production by the evolution 
operator has not been considered earlier.

\section{Entangling and non-entangling operators}

{\bf Definition}. {\it An operator $\hat{A}$ on a Hilbert space $\mathcal{H}$ 
is called entangling if, acting on some disentangled states from this Hilbert 
space $\mathcal{H}$, it produces entangled states, as a result of which its 
measure of entanglement production is nonzero. But when the action of the 
operator $\hat{A}$ on any disentangled state from $\mathcal{H}$ produces  
another disentangled state, so that the operator entanglement production 
measure is zero, such an operator is termed non-entangling}.  

\vskip 2mm 

In order to clearly illustrate how an operator can produce an entangled state
from a disentangled state, let us consider a bipartite system characterized 
by the Hilbert space
\be
\label{11}
 \cH \equiv \cH_1 \bigotimes \cH_2 = {\rm span} \left \{ | n\al \rgl =
| n \rgl \bigotimes | \al \rgl \right \} \;  ,
\ee
composed of two subsystems described by the Hilbert spaces
\be
\label{12}
  \cH_1 = {\rm span} \{ | n \rgl \} \; , \qquad
\cH_2 = {\rm span} \{ | \al \rgl \} \; .
\ee
An operator $\hat{A}$, acting on space (\ref{11}), can be represented
as a resolution
\be
\label{13}
 \hat A =
\sum_{mn} \; \sum_{\al\bt} A_{mn}^{\al\bt} | m\al \rgl \lgl n\bt | \;  ,
\ee
where
$$
A_{mn}^{\al\bt} \equiv \lgl m \al | \hat A | n \bt \rgl \;  .
$$
We assume that the operator possesses a nontrivial trace
$$
 {\rm Tr}_\cH \hat A = \sum_{n\al} A_{nn}^{\al\al} \neq  0 \;  .
$$

The disentangled set consists of disentangled states,
\be
\label{14}
 \cD \equiv \left \{ \vp_{dis} = \vp_1 \bigotimes \vp_2 \right \} \;  .
\ee
In view of the expansions
$$
 \vp_1 = \sum_n a_n | n \rgl \in \cH_1 \; , \qquad
  \vp_2 = \sum_\al b_\al | \al \rgl \in \cH_2 \; ,
$$
the disentangled state can be written as
\be
\label{15}
 \vp_{dis} \equiv \vp_1 \bigotimes \vp_2 =
\sum_{n\al} a_n b_\al | n\al \rgl \;  .
\ee
The action of operator (\ref{13}) on the disentangled state (\ref{15}) gives
\be
\label{16}
\hat A \vp_{dis} =
\sum_{mn} \sum_{\al\bt} A_{mn}^{\al\bt} a_n b_\al | m\al \rgl \; .
\ee
The resulting state (\ref{16}) is entangled if
$$
A_{mn}^{\al\bt} \neq \dlt_{mn} \dlt_{\al\bt} A_n B_\al \;   .
$$
In other words, the operator is entangling, provided it cannot be represented
in the form
$$
\hat A \neq \left ( \sum_n A_n | n \rgl \lgl n | \right ) \bigotimes
\left ( \sum_\al B_\al | \al \rgl \lgl \al | \right ) \; .
$$

For the partial factor operators, we have
$$
 \hat A_1 \equiv {\rm Tr}_{\cH_2} \hat A =
\sum_{mn} \sum_\al A_{mn}^{\al\al} | m \rgl \lgl n | \;  , \qquad
 \hat A_2 \equiv {\rm Tr}_{\cH_1} \hat A =
\sum_n \sum_{\al\bt} A_{nn}^{\al\bt} | \al \rgl \lgl \bt | \;   .
$$
Then the non-entangling counterpart (\ref{5}) becomes
\be
\label{17}
\hat A^{\otimes} =
\frac{\hat A_1 \bigotimes \hat A_2}{{\rm Tr}_\cH \hat A} \;   .
\ee
This yields the measure of entanglement production (\ref{10}), with the norms
$$
|| \hat A_1 ||^2 =
\sum_{mn} \sum_{\al\bt} \left ( A_{mn}^{\al\al} \right )^* A_{mn}^{\bt\bt} \; ,
\qquad
|| \hat A_2 ||^2 =
\sum_{mn} \sum_{\al\bt} A_{mm}^{\al\bt} \left ( A_{nn}^{\al\bt}  \right )^* \; ,
$$
$$
 || \hat A ||^2 =
\sum_{mn} \sum_{\al\bt} \left | A_{mn}^{\al\bt} \right |^2 \;  .
$$

To show by a simple example how an operator can entangle an initially
disentangled state, let us take the operator
\be
\label{18}
 \hat A = C \sum_{mn} | mm \rgl \lgl nn | \;  ,
\ee
where $C$ is a constant. Since
$$
 A_{mn}^{\al\bt} = C \dlt_{m\al} \dlt_{n\bt} \;  ,
$$
we find that the action of this operator on a disentangled state (\ref{15})
results in the state
\be
\label{19}
 \hat A \vp_{dis} = C \left ( \sum_n a_n b_n \right ) \sum_m | mm \rgl \;  .
\ee
This is what is called a multimode state, which is a maximally entangled
state. In the case of only two modes, it represents the well known Bell
state.

To calculate the entanglement production measure in the case of many modes,
we denote their number by $M$, given by the condition
$$
 M \equiv {\rm dim} \cH_1 = {\rm dim} \cH_2 \;  .
$$
Then we have
$$
 || \hat A ||^2 = M^2 | C |^2 \; , \qquad {\rm Tr}_\cH \hat A = MC \;  .
$$
For the partial factor operators
$$
 \hat A_i \equiv {\rm Tr}_{\cH/\cH_i} \hat A  =
C \sum_n | n \rgl \lgl n | \; ,
$$
we get the norms squared
$$
|| \hat A_i ||^2 =| C |^2   \;  .
$$
Therefore the norm squared of the non-entangling counterpart (\ref{17}) is
$$
|| \hat A^{\otimes} ||^2 = \frac {| C |^2}{M^2} \;  .
$$
In this way, we come to the measure of entanglement production,
\be
\label{20}
 \ep(\hat A) = 2 \log M \;  ,
\ee
caused by operator (\ref{18}).

\section{Entangling by evolution operators}

Evolution operators can produce entanglement in the process of natural
system evolution. Suppose, at the initial time $t=0$ the system is prepared
in a disentangled state $\psi(0)$. In the process of its evolution, it
passes to a state $\psi(t)$ that can be entangled by the action of the
evolution operator, since
\be
\label{21}
\psi(t) = \hat U(t) \psi(0) \; , \qquad \hat U(t) = e^{-iHt} \;  ,
\ee
where $H$ is the system Hamiltonian assumed to be independent of time.
Then the produced entanglement can be quantified by the measure of
entanglement production (\ref{8}) or (\ref{10}), with the evolution
operator in the place of $\hat{A}$.

For concreteness, let us take the system Hamiltonian in the form
\be
\label{22}
H = H_1 \bigotimes \hat 1_2 + \hat 1_1 \bigotimes H_2 + H_{int} \;   ,
\ee
characterizing two subsystems with the Hamiltonians $H_1$ and $H_2$,
defined on the Hilbert spaces $\mathcal{H}_1$ and $\mathcal{H}_2$,
respectively, so that
$$
H_1 | n \rgl = E_n | n \rgl \; , \qquad
\cH_1 = {\rm span} \{ | n \rgl \} \; ,
$$
\be
\label{23}
 H_2 | \al \rgl = E_\al | \al \rgl \; , \qquad
\cH_2 = {\rm span} \{ | \al \rgl \} \;  ,
\ee
and interacting by means of an interaction Hamiltonian $H_{int}$. The
notation $\hat{1}_i$ implies a unity operator on the corresponding
space $\mathcal{H}_i$. The system Hamiltonian (\ref{22}) acts on the
Hilbert space (\ref{11}). Let the initial state be disentangled, being
represented as
\be
\label{24}
 \psi(0) = \vp_1 \bigotimes \vp_2 \in \cD \;  .
\ee

The partial evolution operators are
\be
\label{25}
 \hat U_1(t) \equiv {\rm Tr}_{\cH_2} \hat U(t) \; , \qquad
\hat U_2(t) \equiv {\rm Tr}_{\cH_1} \hat U(t) \;  .
\ee
The non-entangling evolution counterpart is of form (\ref{17}), being
\be
\label{26}
 \hat U^{\otimes}(t) =
\frac{\hat U_1(t) \bigotimes \hat U_2(t)}{{\rm Tr}_\cH\hat U(t) } \;  .
\ee
For the measure of entanglement production, we get
\be
\label{27}
 \ep \left ( \hat U(t) \right ) =
\log \; \frac{||\hat U(t)|| }{||\hat U^{\otimes}(t)|| } \equiv \ep(t) \;  .
\ee

The evolution-operator norm is
\be
\label{28}
 || \hat U(t) ||^2 = M_1 M_2 \;  ,
\ee
with the space dimensionalities denoted as
\be
\label{29}
 M_i \equiv {\rm dim} \cH_i \qquad (i = 1,2 ) \;  .
\ee
Thus, measure (\ref{27}) becomes
\be
\label{30}
\ep(t) =
\frac{1}{2} \; \log \; \frac{M_1 M_2}{||\hat U^{\otimes}(t)||^2} \;   .
\ee

At the initial moment of time, before the evolution has started, the
measure of entanglement production has to be zero. To show this, we
need to consider the operators
$$
\hat U(0) = \hat 1_\cH = \hat 1_1 \bigotimes \hat 1_2 \; , \qquad
\hat U_1(0) \equiv {\rm Tr}_{\cH_2} \hat 1_\cH = M_2 \hat 1_1 \; ,
$$
$$
\hat U_2(0) \equiv {\rm Tr}_{\cH_1} \hat 1_\cH = M_1 \hat 1_2 \qquad
 \hat U^{\otimes}(0) =
\frac{\hat U_1(0)\bigotimes\hat U_2(0)}{M_1 M_2} \; .
$$
With the norms squared
$$
|| \hat U_1(0) ||^2 = M_1 M_2^2 \; , \qquad || \hat U_2(0) ||^2 = M_1^2 M_2 \; , \qquad
|| \hat U^\otimes(0) ||^2 = M_1 M_2 \; ,
$$
we find that $\varepsilon(0) = 0$, as it should be.

At finite time, the measure of entanglement production can become non-zero,
which depends on the system Hamiltonian. In some particular cases of the
latter, the evolution operator can be simplified for any finite time
\cite{Bernstein_27,Ramakrishna_28}. For an arbitrary Hamiltonian, one can
consider the short-time behavior. Then, as $t \ra 0$, to second order in $t$,
we have
\be
\label{31}
 \hat U_1(t) \simeq M_2 - it {\rm Tr}_{\cH_2} H - \;
\frac{t^2}{2} \; {\rm Tr}_{\cH_2} H^2  \qquad
 \hat U_2(t) \simeq M_1 - it {\rm Tr}_{\cH_1} H  - \;
\frac{t^2}{2} \; {\rm Tr}_{\cH_2} H^2 \; .
\ee
Introducing the notation
$$
\hat \Delta_1 \equiv M_2 {\rm Tr}_{\cH_2} H^2 -
\left ( {\rm Tr}_{\cH_2} H \right )^2 \; ,
\qquad
\hat \Delta_2 \equiv M_1 {\rm Tr}_{\cH_1} H^2 -
\left ( {\rm Tr}_{\cH_1} H \right )^2 \; ,
$$
\be
\label{32}
 \hat \Delta_{12} \equiv M_1 M_2 {\rm Tr}_\cH H^2 -
\left ( {\rm Tr}_\cH H \right )^2 \;  ,
\ee
we find
$$
|| \hat U_1(t) ||^2 \simeq M_1 M_2^2 -
\left ( {\rm Tr}_{\cH_1} \hat \Delta_1 \right ) t^2 \; ,
\qquad
|| \hat U_2(t) ||^2 \simeq M_1^2 M_2 -
\left ( {\rm Tr}_{\cH_2} \hat \Delta_2 \right ) t^2 \; ,
$$
\be
\label{33}
 | {\rm Tr}_\cH \hat U(t) |^2 \simeq M_1^2 M_2^2 -  \Delta_{12}  t^2 \;   .
\ee
Therefore
\be
\label{34}
 || \hat U^\otimes(t) ||^2 \simeq M_1 M_2 - \mu t^2 \;  ,
\ee
where
$$
\mu \equiv \frac{1}{M_1M_2} \left ( M_1 {\rm Tr}_{\cH_1} \hat\Delta_1 +
 M_2 {\rm Tr}_{\cH_2} \hat\Delta_2 - \Delta_{12} \right ) \;  .
$$
Finally, we obtain the short-time behavior of the entanglement-production
measure
\be
\label{35}
 \ep(t) \simeq \frac{1}{2} \; \mu t^2 \qquad ( t \ra 0 ) \;  ,
\ee
calculated to second order in $t$. Here, we keep in mind the natural logarithm 
in definition (\ref{27}). Dealing with the logarithm over the base $2$, we 
should replace $\mu$ by $\mu/\ln2$.

At the initial stage, the entanglement production is quadratic in time.

\section{Heisenberg evolutional entanglement}

As an illustration, let us consider a bipartite system characterized 
by spins ${\bf S}_j = \{ S_j^\alpha \}$, with the Heisenberg interaction. 
Such spin ensembles represent many finite-state systems widely studied 
in a variety of physics applications as well as in information processing. 
The Hamiltonian is a sum of two terms:
\be
\label{36}
  H= H_0 + H_{int} \; ,
\ee
where the first term has the Zeeman structure
\be
\label{37}
 H_0 = - h \left ( S_1^z \bigotimes \hat 1_2 +
\hat 1_1 \bigotimes S_2^z \right ) \;  ,
\ee
and the second term describes an anisotropic Heisenberg interaction
\be
\label{38}
 H_{int} = J_1 \left ( S_1^x \bigotimes S_2^x + S_1^y \bigotimes S_2^y
\right )  + 2J \; S_1^z \bigotimes S_2^z \;  .
\ee
The Heisenberg model is defined for any dimensionality of spins $\bf{S}_j$. 
Here, we shall consider spins one-half, with the standard relation of spin
components with the Pauli matrices: $S_j^\alpha = (1/2) \sigma_j^\alpha$.  

Using the ladder operators $S_j^{\pm} \equiv S_j^x \pm S_j^y$ reduces the 
interaction term to the form
\be
\label{39}
 H_{int} = 2J\;  S_1^z \bigotimes S_2^z + J_1 \left (
 S_1^+ \bigotimes S_2^- +  S_1^- \bigotimes S_2^+ \right ) \;  .
\ee
The interaction parameters $J$ and $J_1$ can be of any sign.

Considering the entanglement production by the evolution operator, we follow
the previous sections, omitting the details of the calculational procedure
that is delineated in the Appendix A. For expressions (\ref{32}), we find
$$
\hat\Delta_1 = \left ( h^2 + J^2 + 2J_1^2 \right ) \hat 1_1 -
4J h S_1^z \; ,
\qquad
\hat\Delta_2 = \left ( h^2 + J^2 + 2J_1^2 \right ) \hat 1_2 -
4J h S_2^z \; ,
$$
$$
\Delta_{12} = 4 \left ( 2h^2 + J^2 + 2J_1^2 \right ) \; .
$$
The norm of the non-entangling evolution-operator counterpart  (\ref{26}), 
at short time, reads as
$$
|| \hat U^\otimes(t) ||^2 \simeq
8\;\frac{2-(h^2+J^2+2J_1^2)t^2}{4-(2h^2+J^2+2J_1^2)t^2} \;   .
$$
Then, for the entanglement production measure (\ref{27}) at the initial stage,
we obtain
\be
\label{40}
 \ep(t) \simeq \frac{1}{8} \left ( J^2 + 2J_1^2
\right ) t^2 \qquad (t\ra 0) \;  ,
\ee
in agreement with the quadratic in time behavior (\ref{35}).

Note that at this initial stage, the evolutional entanglement is produced
by spin interactions, while an external field is not yet playing role.

\section{Ising evolutional entanglement}

In order to analyze the behavior of the entanglement-production measure
for all times, let us consider a system with strongly anisotropic
Heisenberg interactions yielding the Ising Hamiltonian
$$
H = H_0 + H_{int} \; ,
$$
\be
\label{41}
  H_0 = - h \left ( S_1^z \bigotimes \hat 1_2 + \hat 1_1 \bigotimes S_2^z
\right ) \; , \qquad  H_{int} = 2J\; S_1^z \bigotimes S_2^z \; .
\ee
In view of the commutator
$$
[ H_0 , \; H_{int} ] = 0 \;   ,
$$
we have
\be
\label{42}
 e^{-i H t} = e^{-iH_0 t}  e^{-i H_{int} t} \;  .
\ee

Expanding the exponents in Taylor series and summing back, as is explained
in the Appendix B, we find for the exponent with the Zeeman term
\be
\label{43}
e^{-iH_0 t} = 1 + \frac{H_0^2}{h^2} \; [ \cos(ht) -1 ] - i\;
\frac{H_0}{h} \; \sin(ht) \;  ,
\ee
while for the exponent with the interaction term, we get
\be
\label{44}
 e^{-iH_{int} t} = \cos \left ( \frac{Jt}{2} \right ) -
2i\; \frac{H_{int}}{J}\; \sin  \left ( \frac{Jt}{2} \right ) \; .
\ee
Then the evolution operator can be represented as
$$
 e^{-iH t} = \left \{ 1 + \frac{H_0^2}{h^2} \; [ \cos(ht) -1 ]\right \}
\cos \left ( \frac{Jt}{2} \right ) - \;
\frac{H_0}{h}\; \sin(ht) \sin  \left ( \frac{Jt}{2} \right ) -
$$
\be
\label{45}
 - i \left \{\frac{2H_{int}}{J} + \frac{H_0^2}{h^2} \; [ \cos(ht) -1 ] \right\}
\sin  \left ( \frac{Jt}{2} \right ) -
i \; \frac{H_0}{h} \; \sin(ht) \cos\left ( \frac{Jt}{2} \right ) .
\ee

For the partially-traced operators, defined in Eqs. (\ref{25}), we have
$$
 \hat U_j(t) = [ 1 + \cos(ht) ] \cos\left ( \frac{Jt}{2} \right ) \hat 1_j +
2S_j^z \sin(ht) \sin\left ( \frac{Jt}{2} \right ) +
$$
\be
\label{46}
 + i  [ 1 - \cos(ht) ] \sin\left ( \frac{Jt}{2} \right ) \hat 1_j +
2i S_j^z \sin(ht) \cos\left ( \frac{Jt}{2} \right ) \; ,
\ee
where $j = 1,2$. Their norms squared  are given by the formula
\be
\label{47}
 || \hat U_j(t)||^2 = 4 [ 1 + \cos(ht)\cos(Jt) ] \;  .
\ee
And for the evolution operator, we obtain the trace
\be
\label{48}
{\rm Tr}_\cH \hat U(t) = 2 [ 1 + \cos(ht) ] \cos\left ( \frac{Jt}{2} \right )
+2i [ 1 - \cos(ht) ] \sin\left ( \frac{Jt}{2} \right ) \; ,
\ee
which yields
\be
\label{49}
 | {\rm Tr}_\cH \hat U(t) |^2 = 4 [ 1 + \cos^2(ht) + 2\cos(ht)\cos(JT) ] \;  .
\ee

Finally, the entanglement-production measure (\ref{27}), caused by the
evolution operator, is
\be
\label{50}
\ep(t) = \log \;
\frac{\sqrt{1+\cos^2(ht)+2\cos(ht)\cos(Jt)} }{1+\cos(ht)\cos(Jt)} \;  .
\ee
This measure, as is straightforward to check, is positive for all times
$t > 0$. It tends to zero at the beginning of the evolution as
\be
\label{51}
\ep(t) \simeq \frac{J^2}{8\ln 2}\; t^2 +
\frac{J^2(J^2-12h^2)}{192\ln 2} \; t^4 \;   ,
\ee
as it should be according to Eq. (\ref{35}). Here we use the logarithm
over the base $2$. Again, we see that the first term does not depend on
the field $h$ that enters only in the higher terms.

The measure does not depend on the signs of $h$ and $J$. But the existence
of interactions is crucial, since without interactions there is no
entanglement at all:
\be
\label{52}
\lim_{J\ra 0} \ep(t) = 0 \; .
\ee

The existence of the field $h$ is also important. This is due to the
invariance of Hamiltonian (\ref{41}) with respect to the spin inversion
$S_j^z \ra - S_j^z$, when $h \equiv 0$, while this invariance is absent
for any finite $h$. In the case of this invariance, when $h \equiv 0$,
we have
$$
\ep(t) = \frac{1}{2} \; \log\; \frac{2}{1+\cos(Jt)} \qquad
( h \equiv 0 ) \;   .
$$
This expression diverges at the moments of time $\pi(1+2n)/J$, where
$n = 0,1,2,\ldots$. On the contrary, at these moments of time,
$\varepsilon(t)$, defined by Eq. (\ref{50}), is zero, when $h \neq 0$
is any finite quantity, except special points of the set of zero
measure to be defined below. The singularity points correspond to the
exceptional conditions when either
\be
\label{53}
 \frac{h}{J} = \frac{2p}{1+2n} \; , \qquad
t_1 = (1+2n) \; \frac{\pi}{J} \;  ,
\ee
or when
\be
\label{54}
 \frac{h}{J} = \frac{1+2n}{2p} \; , \qquad
t_2 = 2p \; \frac{\pi}{J} \;  ,
\ee
with $n = 0,1,2,\ldots$ and $p = 1,2,\ldots$. For all other $h$, there are
no singularities.

Generally, the entanglement production measure (\ref{50}) is quasi-periodic,
with the periods
\be
\label{55}
 T_1 = \frac{\pi}{|h|}\; , \qquad   T_2 = \frac{2\pi}{|h+J|}\; , \qquad
 T_3 = \frac{2\pi}{|h-J|}\; ,
\ee
except when the periods are commensurable. Thus, when $h/J$ is an irreducible
rational number $h/J = p/q$, where $p$ and $q$ both are odd numbers, then
expression (\ref{50}) is periodic, with the period $T = \pi q$. And when
$h/J$ is rational, such that $h/J = p/q$, where one of the integers is even,
while the other is odd, then function (\ref{50}) is periodic, with the
period $T = 2\pi q$.

The typical temporal behavior of measure (\ref{50}), as a function of
time measured in units of $1/J$, is shown in Figs. 1 and 2. The logarithm
is taken with respect to base $2$. The field $h$ is measured in units
of $J$. Figure 1 shows the cases of periodic behavior, while Fig. 2
illustrates quasi-periodic entanglement-production.

By changing the system parameters, it is possible to regulate the 
evolutional process of entanglement production.

\section{Conclusion}

When a system is in a disentangled state but one needs to transfer it into 
an entangled state, two ways are possible, which can be classified as 
external and internal. One way is when entanglement is generated in a system 
by resorting to externally imposed appropriate transformations. Some of the 
related cases have been considered earlier. For example, by an external 
alternating field it is possible to generate multiple entangled modes in a
Bose-Einstein condensate. The other possibility is to allow for the system 
to naturally evolve according to the evolution law prescribed by the evolution 
operator. It is this second way that is studied in the present paper. The 
entanglement production generated by the evolution operator has not been 
considered in previous literature.    

Entanglement production, generated by an evolution operator $\hat{U}(t)$,
and quantified by the entanglement production measure $\varepsilon(\hat{U}(t))$,
is investigated. As illustrations, we consider the bipartite systems with spin
interactions of the Heisenberg and Ising types. Such spin objects are typical
for many finite-level or finite-state physical systems that can be employed
for information processing. The measure of entanglement production oscillates
in time, being in general quasi-periodic. The existence of interactions is
crucial for this measure to be nonzero. Without interactions, no entanglement
is produced.

The evolutional entanglement, produced by the evolution operator, as studied
in the present paper, is different from the entanglement produced by a
time-dependent statistical operator $\hat{\rho}(t)$ of a nonequilibrium system,
as considered in Refs. \cite{Yukalov_10,Yukalov_11,Yukalov_12,Yukalov_17}. The
entanglement production measure, analyzed in the latter papers, has been
$$
 \ep(\hat\rho(t)) =
\log\; \frac{||\hat\rho(t)||}{||\hat\rho^\otimes(t)||} \;  ,
$$
where, according to the general definition (\ref{8}), the disentangled, or
distilled, statistical operator is
$$
 \hat\rho^\otimes(t) = \bigotimes_{i=1}^N \hat\rho_i(t) \;  ,
$$
with the partial operators
$$
 \hat\rho_i(t) = {\rm Tr}_{\cH/\cH_i} \hat\rho(t) \;  .
$$

Also, the entanglement production, generated by quantum operations, should
not be confused with the state entanglement that is quantified by other
measures \cite{Williams_1,Nielsen_2,Vedral_3,Keyl_4,Wilde_5}. For example,
the entanglement of a state, corresponding to a statistical operator
$\hat{\rho}(t)$, can be quantified by the relative entropy
$$
D(t) = {\rm Tr}_\cH \; \hat\rho(t) \;
\ln\; \frac{\hat\rho(t)}{\hat\rho^\otimes(t)} \;   ,
$$
that is also called the Kullback-Leibler distance, since it shows the
distance of the state $\hat{\rho}(t)$ from the distilled state
$\hat{\rho}^{\otimes}(t)$. When several distillations are admissible, one
considers the minimum of the above distance.

By studying the entanglement production, caused by the evolution operator,
it is possible to evaluate the period of time during which the considered 
system would evolve from an initial disentangled state to an entangled state. 
This method of following the natural evolution of the system provides an 
alternative to the procedure of creating entanglement by means of external 
transformations.

\vskip 5mm

{\bf Acknowledgement}. Financial support from RFBR (grant $\#$14-02-00723)
is appreciated.

\newpage

{\parindent=0pt
{\large {\bf Appendix A}}. Evolutional entanglement production for Heisenberg
interactions }

\vskip 5mm
Calculating the operator norms, we meet the powers of the Hamiltonian, which
are represented in the standard symmetrized form. For instance, the squares
of the sums of two operators are given by the expressions
$$
\left (\hat A_i + \hat B_i \right )^2 = \hat A_i^2 + \hat B_i^2 +
\hat A_i \hat B_i + \hat B_i \hat A_i \; ,
$$
$$
 \left (\hat A_1 \bigotimes \hat A_2  + \hat B_1 \bigotimes \hat B_2 \right )^2
= \hat A_1^2 \bigotimes \hat A_2^2 + \hat B_1^2 \bigotimes \hat B_2^2 +
\hat A_1 \hat B_1 \bigotimes \hat A_2 \hat B_2 +
\hat B_1 \hat A_1 \bigotimes  \hat B_2 \hat A_2 \; .
$$
In that way, for Hamiltonians (\ref{37}) and (\ref{38}), we have
$$
H_0^2 = \frac{h^2}{2} \left ( \hat 1_\cH + 4S_1^z \bigotimes S_2^z \right ) \; ,
$$
$$
H_{int}^2 = \frac{1}{4} \left ( J^2 + 2J_1 \right ) \hat 1_\cH -
 2 J_1^2 \; S_1^z \bigotimes S_2^z
 - J J_1 \left (  S_1^+ \bigotimes  S_2^- +  S_1^- \bigotimes  S_2^+  \right ) \;  ,
$$
$$
H_0 H_{int} = H_{int} H_0 = \frac{1}{2} \; J H_0 \; .
$$
Then for Hamiltonian (\ref{36}), we get
$$
H^2 = H_0^2 + JH_0 + H_{int}^2 \;   .
$$
Under spins one-half, the basis can be taken as a set of two vectors,
corresponding to spin up and spin down. So that $M_1 = M_2 = 2$.

The following traces are found:
$$
{\rm Tr}_{\cH_1} H_0 = - 2h S_2^z \; , \qquad
{\rm Tr}_{\cH_2} H_0 = - 2h S_1^z \; , \qquad
{\rm Tr}_{\cH_1} H_{int}={\rm Tr}_{\cH_2} H_{int} = 0 \; ,
$$
$$
 {\rm Tr}_{\cH_1} H_0^2  = h^2 \hat 1_2 \; , \qquad
{\rm Tr}_{\cH_2} H_0^2  = h^2 \hat 1_1 \;  ,
$$
$$
 {\rm Tr}_{\cH_1} H_{int}^2 = \left ( \frac{1}{2} \; J^2 + J_1^2 \right ) \hat 1_2 \; , \qquad
 {\rm Tr}_{\cH_2} H_{int}^2 = \left ( \frac{1}{2} \; J^2 + J_1^2 \right ) \hat 1_1 \; ,
$$
$$
 {\rm Tr}_{\cH_1} H = -2h S_2^z \; , \qquad {\rm Tr}_{\cH_2} H = -2h S_1^z \; ,
$$
$$
{\rm Tr}_{\cH_1} H^2 = \left ( h^2 + \frac{1}{2} \; J^2 + J_1^2 \right ) \hat 1_2 -
2J h S_2^z \; , \qquad
{\rm Tr}_{\cH_2} H^2 = \left ( h^2 + \frac{1}{2} \; J^2 + J_1^2 \right ) \hat 1_1 -
2J h S_1^z \; ,
$$
$$
{\rm Tr}_\cH H = 0 \; , \qquad {\rm Tr}_\cH H^2 = 2h^2 + J^2 + 2J_1^2 \; .
$$
This results in measure (\ref{40}).

\newpage

{\parindent=0pt
{\large {\bf Appendix B}}. Evolutional entanglement production for Ising
interactions }

\vskip 5mm
Expanding the exponent $\exp(- i H_0 t)$, we meet the terms
$$
H_0^3 = h^2 H_0 \; , \qquad H_0^4 = h^2 H_0^2 \; , \qquad
H_0^5 = h^4 H_0 \; , \qquad H_0^6 = h^4 H_0^2 \; ,
$$
and so on, resulting in the relations
$$
 H_0^{2n} = h^{2(n-1)} H_0^2 \; , \qquad H_0^{2n+1} = h^{2n} H_0 \;  ,
$$
which lead to Eq. (\ref{43}).

Expanding the exponent $\exp(- i H_{int} t)$, we find
$$
 H_{int}^2 = \left ( \frac{J}{2} \right )^2 \; , \qquad
H_{int}^3 = \left ( \frac{J}{2} \right )^2 H_{int}\; , \qquad
 H_{int}^4 = \left ( \frac{J}{2} \right )^4 \; , \qquad
H_{int}^5 = \left ( \frac{J}{2} \right )^4 H_{int}\; ,
$$
and so on, which gives the equations
$$
H_{int}^{2n} = \left ( \frac{J}{2} \right )^{2n} \; , \qquad
H_{int}^{2n+1} = \left ( \frac{J}{2} \right )^{2n} H_{int}\;   .
$$
These relations yield Eq. (\ref{44}).

\newpage

\begin{center}
{\Large{\bf Figure Captions }}

\end{center}

\vskip 3cm

{\bf Figure 1}. The entanglement-production measure, for the case of periodic
evolution, as a function of time measured in units of $1/J$, for different
fields: (a) $h/J = 1$ (the period is $\pi$); (b) $h/J = 5/7$ (the period
is $7\pi$); (c) $h/J = 7$ (the period is $\pi$); (d) $h/J = 8$ (the period
is $2\pi$).

\vskip 1cm
{\bf Figure 2}. The measure of evolutional entanglement production, illustrating
quasi-periodic behavior, for different fields: (a) $h/J = \sqrt{2}$; (b) $h/J =\sqrt{3}/2$; 
(c) $h/J = \sqrt{5}$; (d) $h/J = \sqrt{7}$.

\newpage

\begin{figure}[ht]
\centerline{
\hbox{ \includegraphics[width=7.5cm]{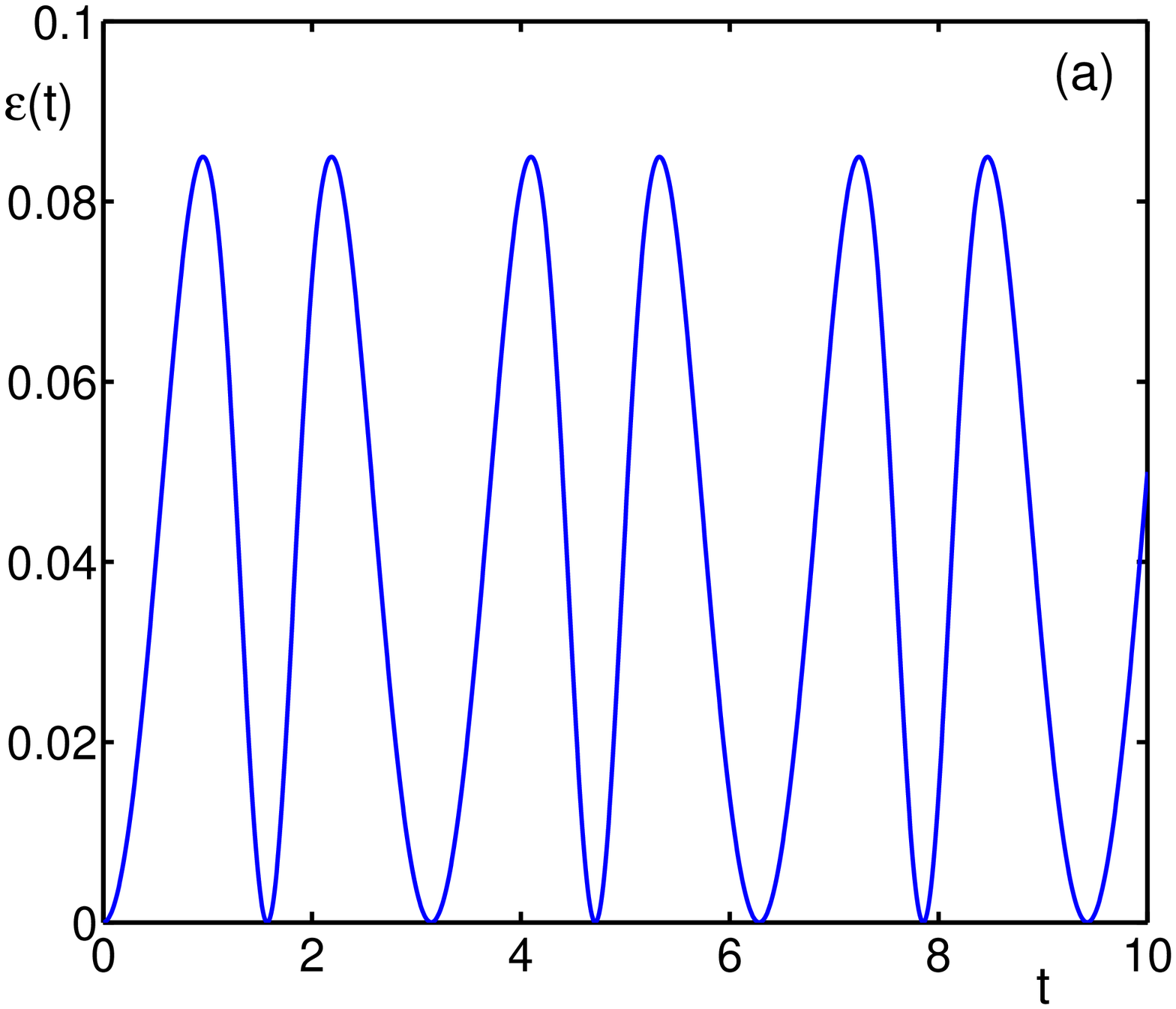} \hspace{1cm}
\includegraphics[width=7.5cm]{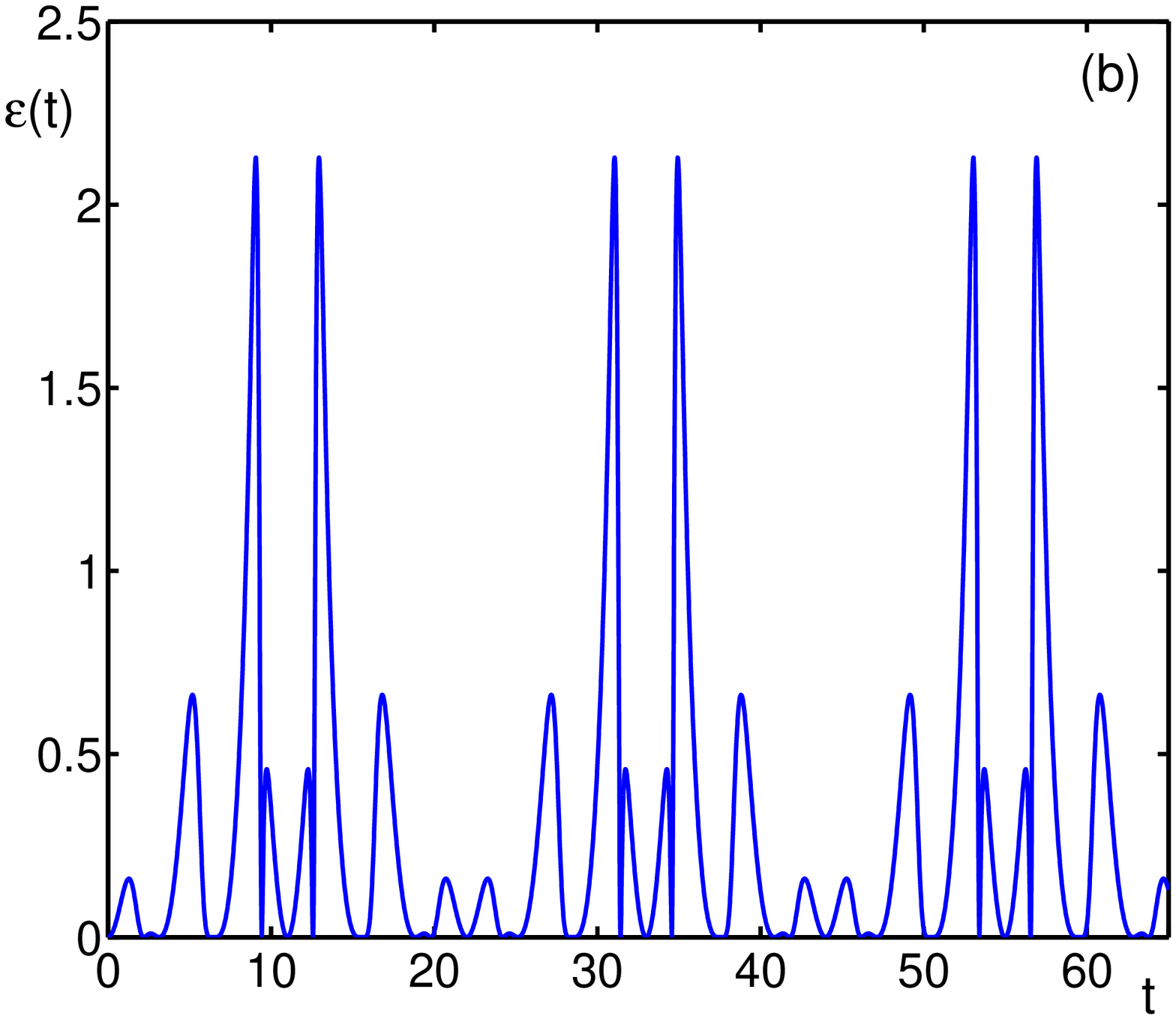}  } }
\vspace{12pt}
\centerline{
\hbox{ \includegraphics[width=7.5cm]{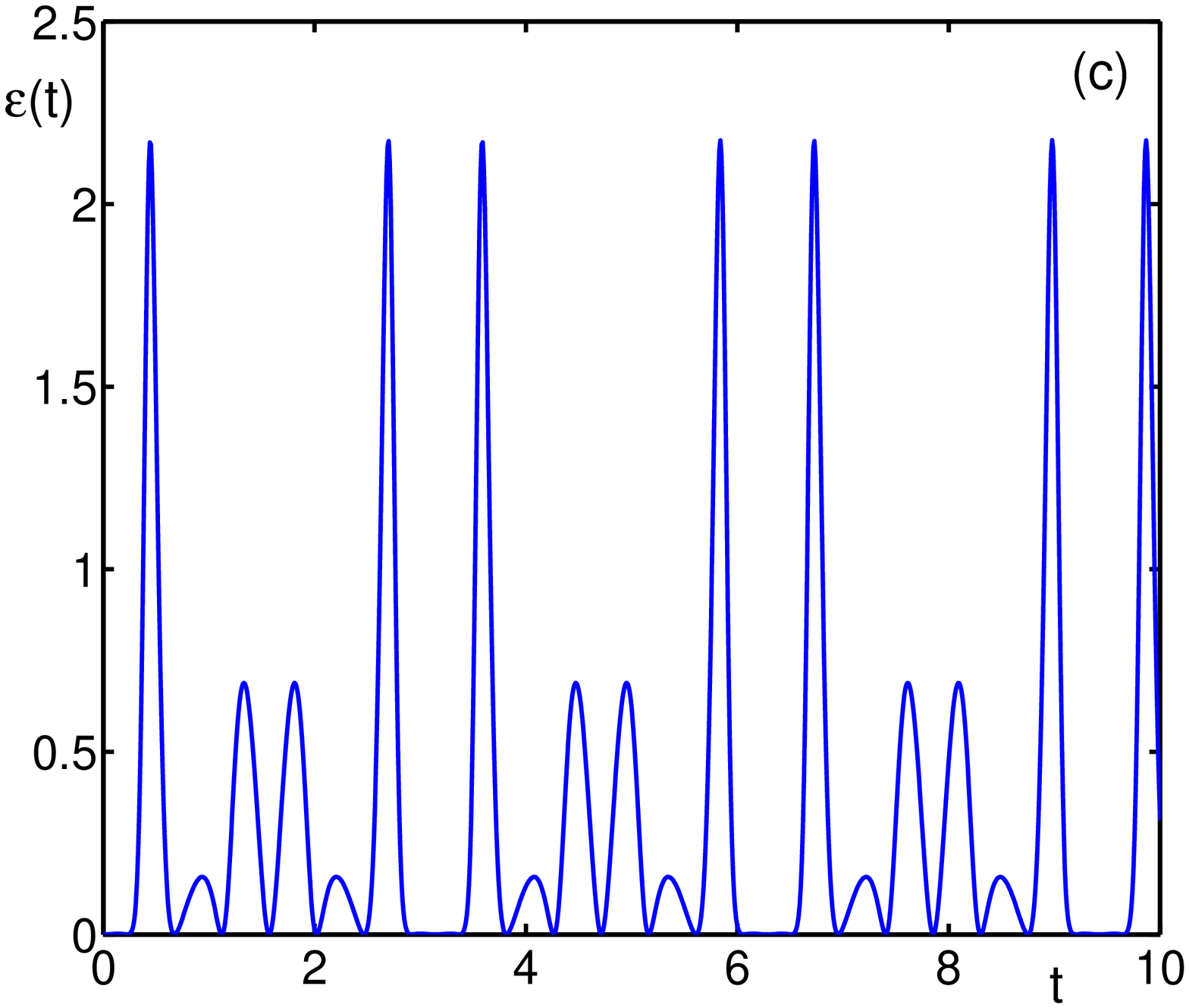} \hspace{1cm}
\includegraphics[width=7.5cm]{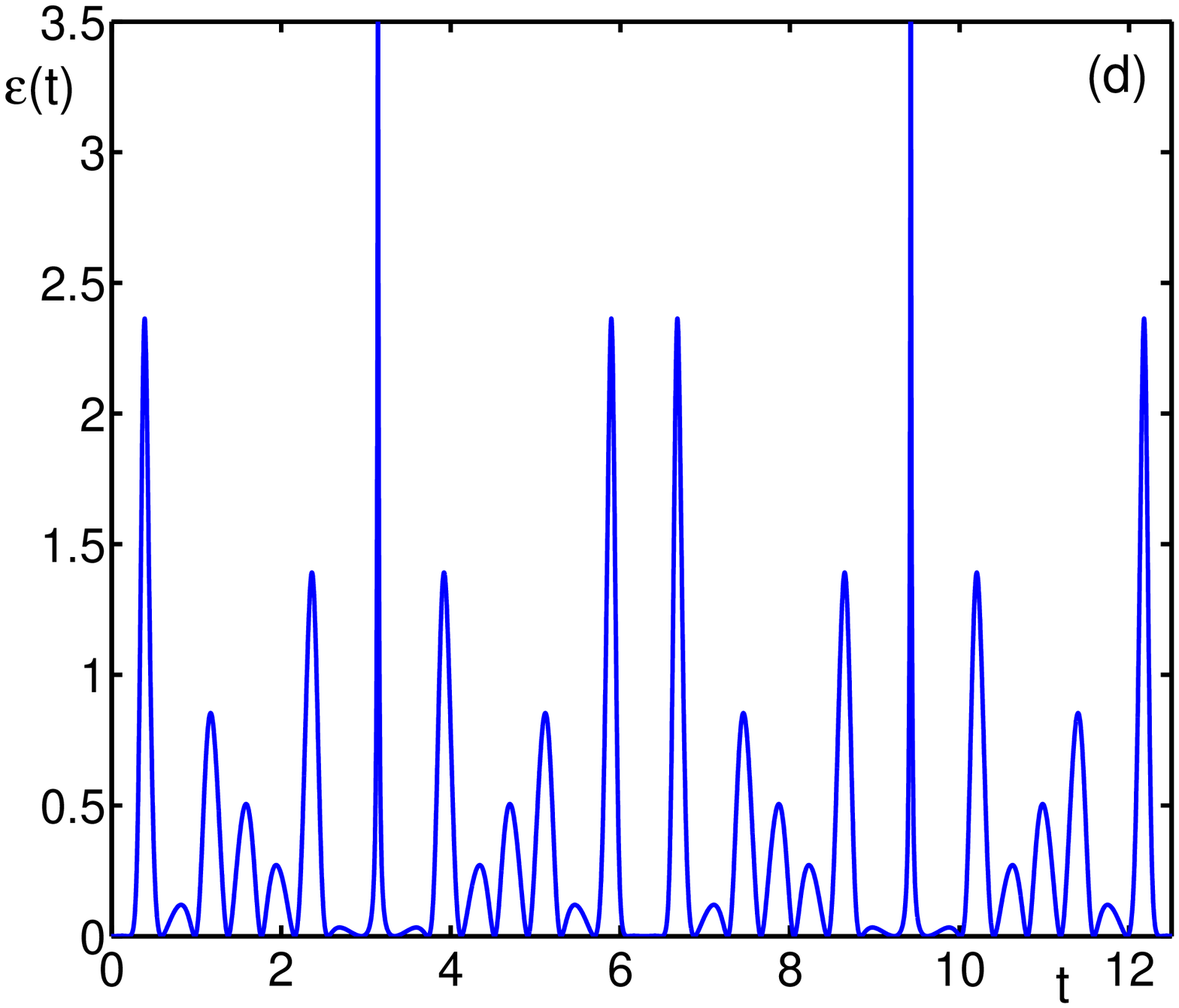} } }
\caption{The entanglement-production measure, for the case of periodic
evolution, as a function of time measured in units of $1/J$, for different
fields: (a) $h/J = 1$ (the period is $\pi$); (b) $h/J = 5/7$ (the period
is $7\pi$); (c) $h/J = 7$ (the period is $\pi$); (d) $h/J = 8$ (the period
is $2\pi$). 
}
\label{fig:Fig.1}
\end{figure}

\newpage

\begin{figure}[ht]
\centerline{
\hbox{ \includegraphics[width=7.5cm]{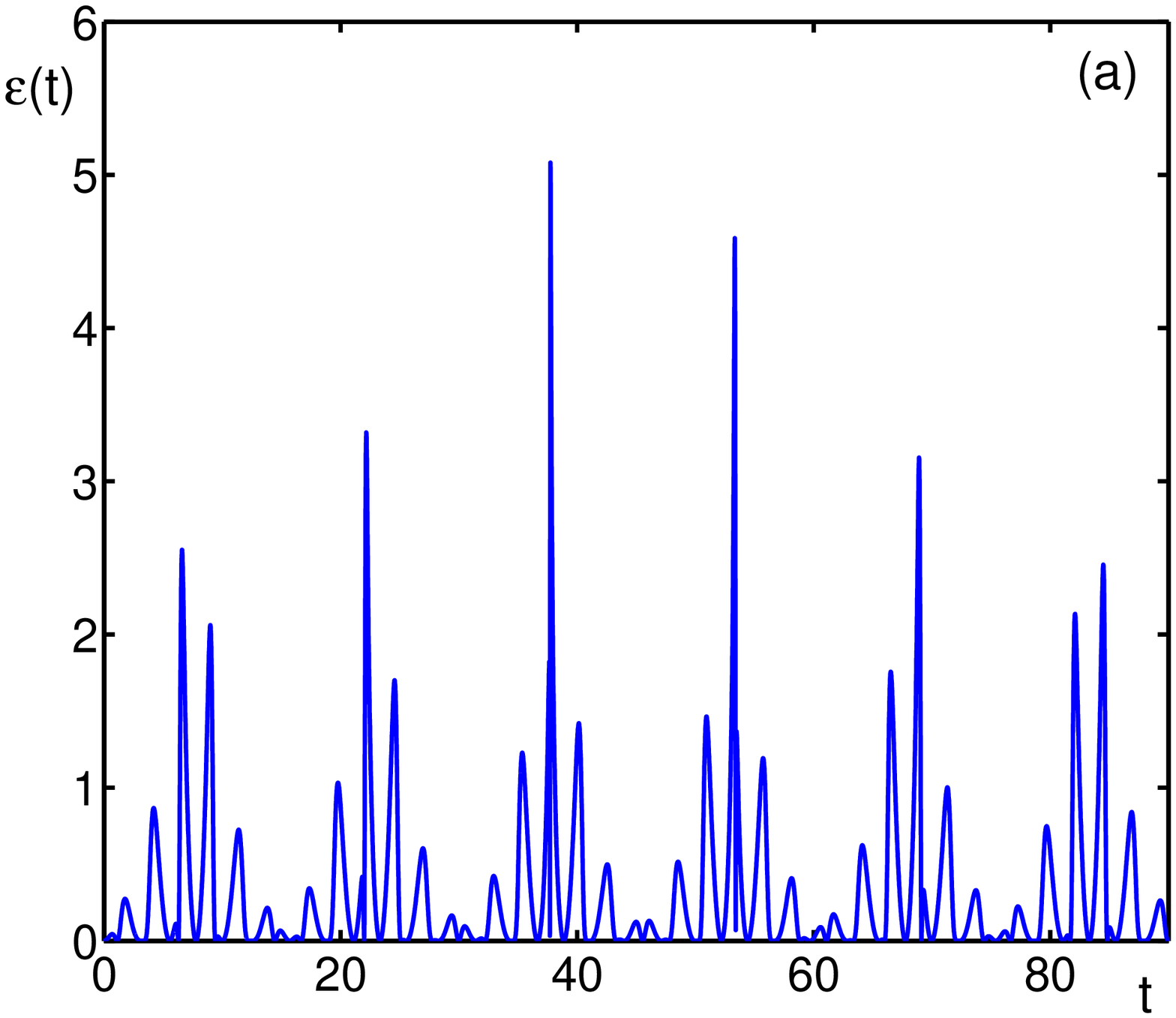} \hspace{1cm}
\includegraphics[width=7.5cm]{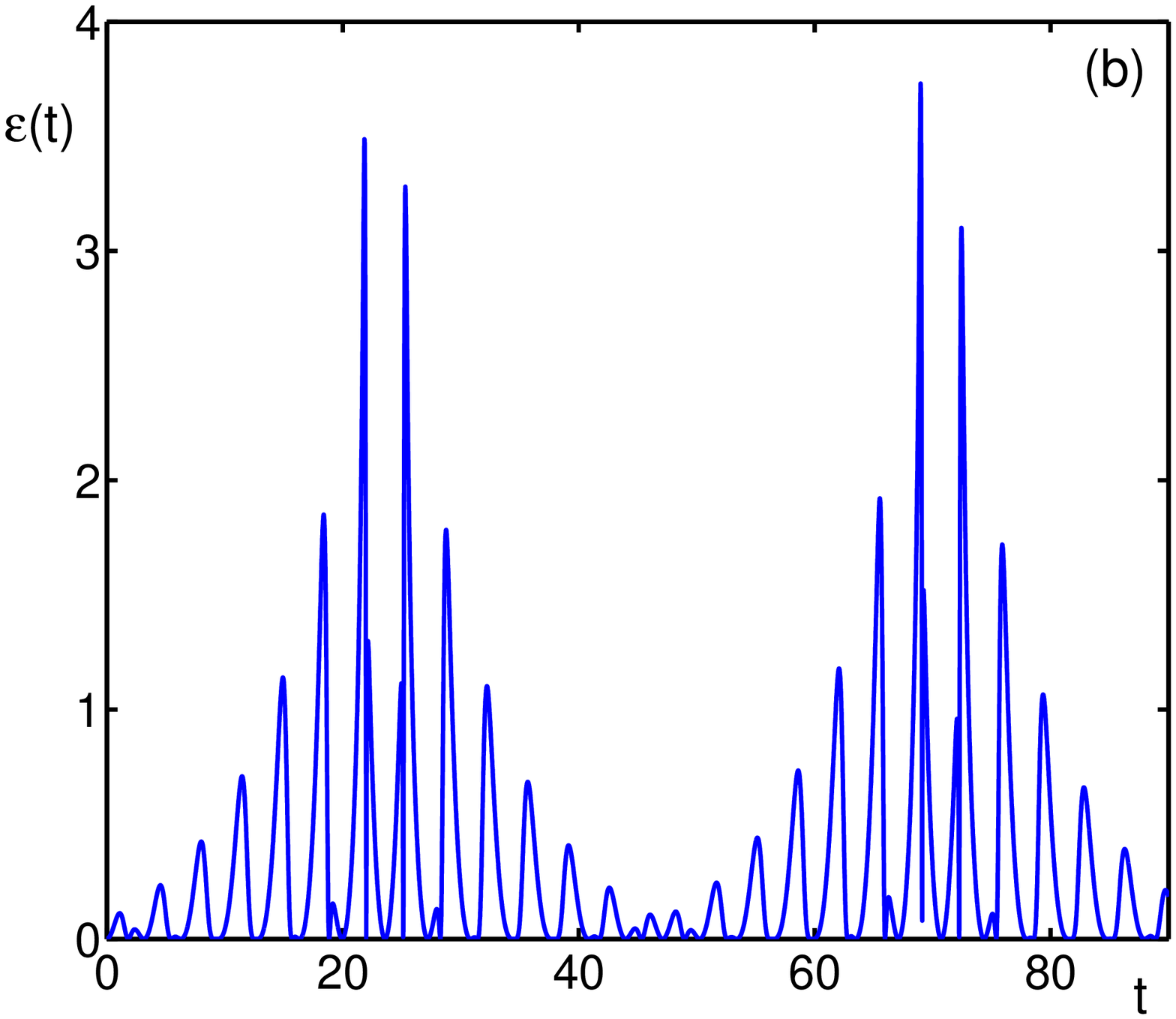}  } }
\vspace{12pt}
\centerline{
\hbox{ \includegraphics[width=7.5cm]{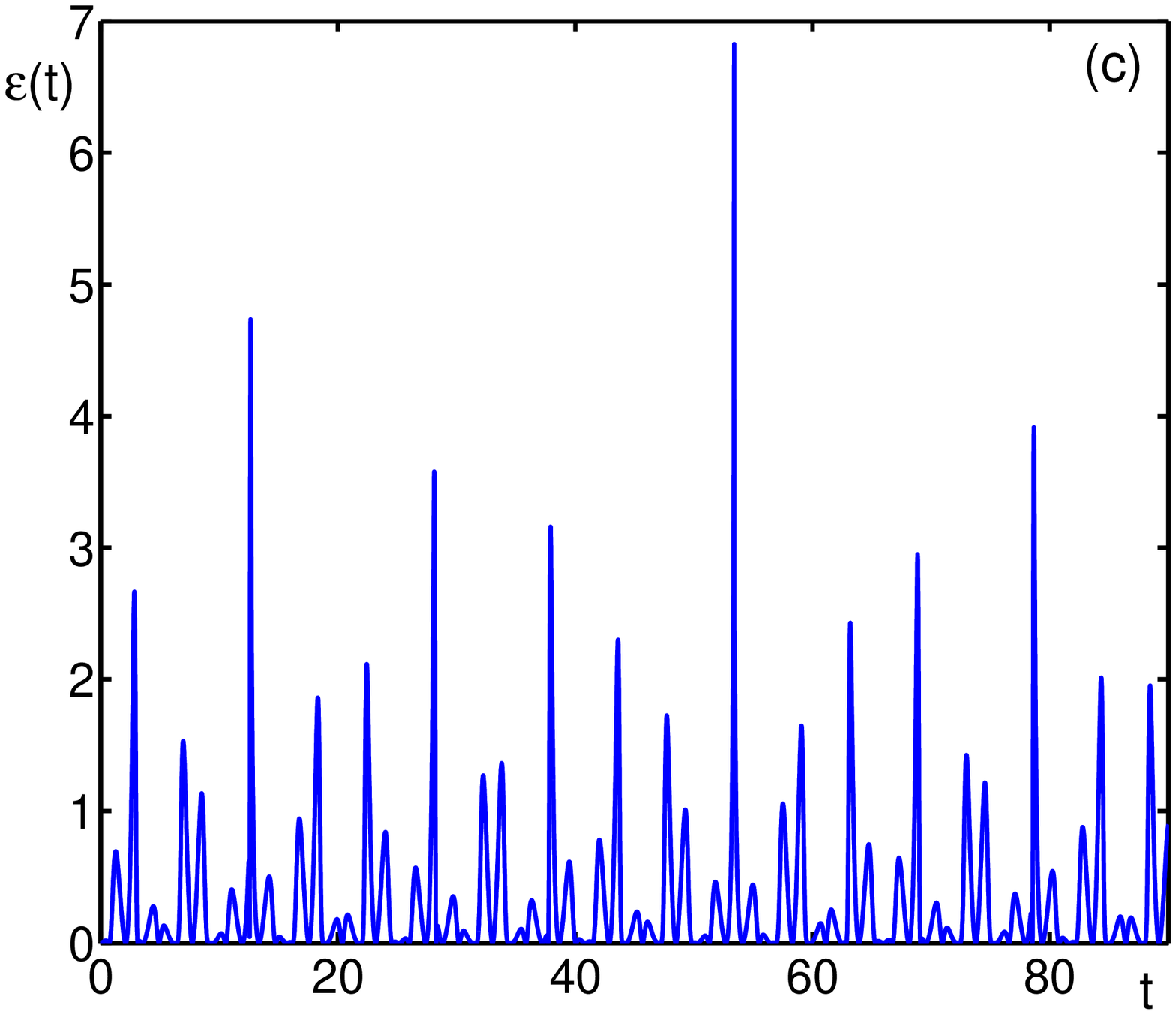} \hspace{1cm}
\includegraphics[width=7.5cm]{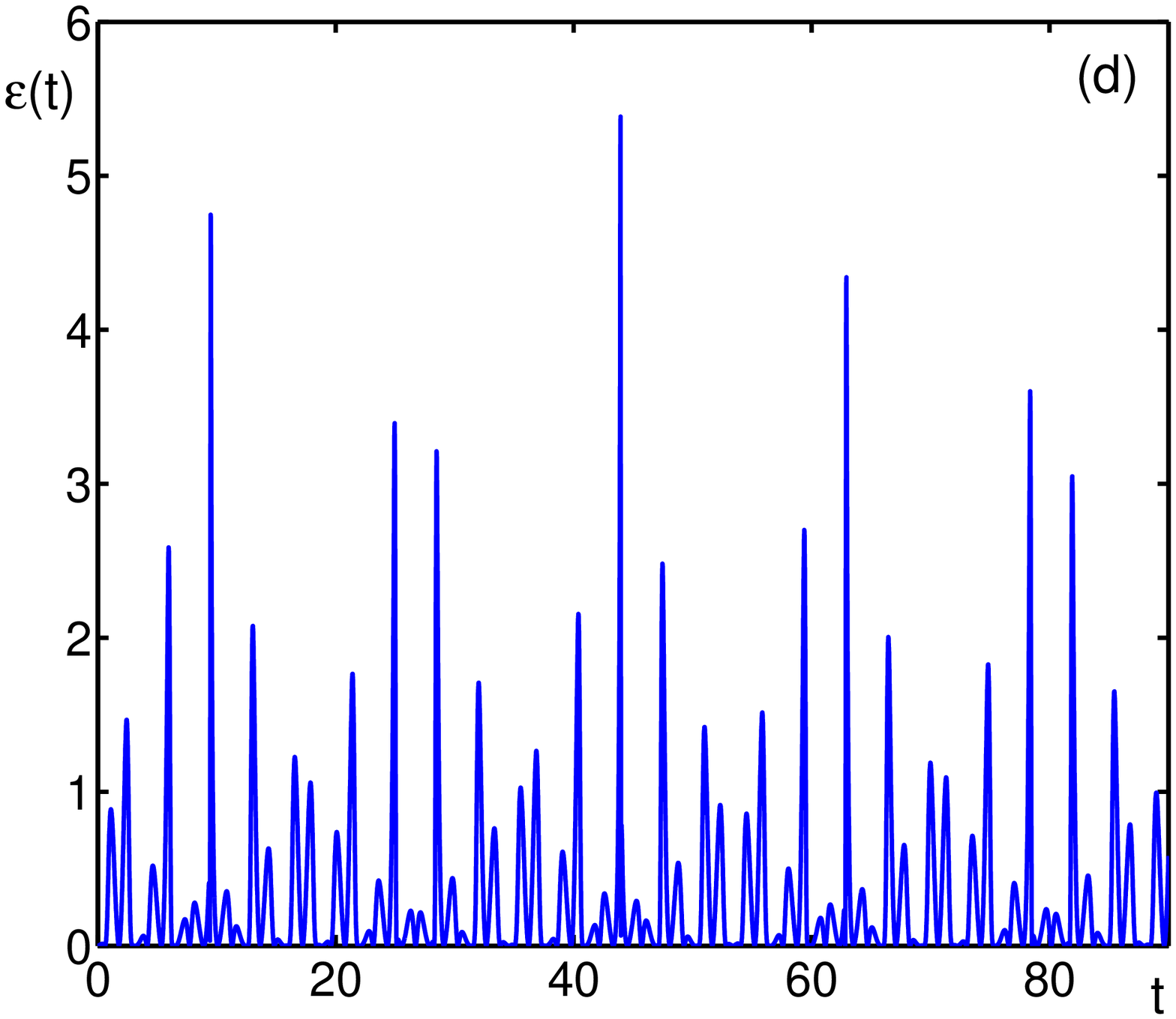} } }
\caption{The measure of evolutional entanglement production, illustrating
quasi-periodic behavior, for different fields: (a) $h/J = \sqrt{2}$;
(b) $h/J = \sqrt{3}/2$; (c) $h/J = \sqrt{5}$; (d) $h/J = \sqrt{7}$.
}
\label{fig:Fig.2}
\end{figure}

\end{document}